# Observation of Superfluidity of Polaritons in Semiconductor Microcavities

A. Amo[1*], J. Lefrère[1], S. Pigeon[2], C. Adrados[1], C. Ciuti[2], I. Carusotto[3‡], R. Houdré[4], E. Giacobino[1], and A. Bramati[1*]

[1]*Laboratoire Kastler Brossel, Université Paris 6, Ecole Normale Supérieure et CNRS, UPMC Case 74, 4 place Jussieu, 75252 Paris Cedex 05, France.*

[2] *Laboratoire Matériaux et Phénomènes Quantiques, UMR 7162, Université Paris Diderot-Paris 7, 75013 Paris, France*

[3] *BEC-CNR-INFM and Dipartimento di Fisica. Universita di Trento. I-38050 Povo, Italy.*

[4] *Institut de Photonique et d'Electronique Quantique, Ecole Polytechnique Fédérale de Lausanne (EPFL), Station 3, CH-1015 Lausanne, Switzerland.*

[‡] *Also at Institute of Quantum Electronics, ETH Zurich, 8093 Zurich, Switzerland.*

[*]e-mail: alberto.amo@spectro.jussieu.fr; bramati@spectro.jussieu.fr

**One of the most striking manifestations of quantum coherence in interacting boson systems is superfluidity. Exciton-polaritons in semiconductor microcavities are two-dimensional composite bosons predicted to behave as particular quantum fluids. We report the observation of superfluid motion of polaritons created by a laser in a semiconductor microcavity. Superfluidity is investigated in terms of the Landau criterion and manifests itself as the suppression of scattering from defects when the flow velocity is slower than the speed of sound in the fluid. On the other hand, a Cerenkov-like wake pattern is clearly observed when the flow velocity exceeds the speed of sound. The experimental findings are in excellent quantitative agreement with the**

**predictions based on a generalized Gross-Pitaevskii theory, showing that polaritons in microcavities constitute a very rich system for exploring the physics of non-equilibrium quantum fluids.**

In a quantum degenerate gas of bosons, the phenomenon of superfluidity can occur as a result of particle-particle interactions, with the spectacular consequence that the quantum fluid is able to flow without friction. Since its first discovery in liquid 4-Helium[1, 2], superfluidity has stimulated the development of a quantum theory of many body systems, and has been observed in several other systems. Most remarkable, recent experiments have investigated the subtle links between superfluidity and Bose-Einstein condensation in gases of ultracold trapped atoms[3-5]. In the past few years, it has been anticipated that exciton-polaritons could also be driven into a quantum fluid regime[6-8], thus stimulating efforts to experimentally demonstrate superfluidity. However, in spite of some indirect observations such as pinned quantized vortices[9] and Bogoliubov-like dispersions[10], and strong indications given by pioneering experiments in polariton parametric oscillators[11], a direct demonstration of exciton-polaritons superfluidity is still missing. In this work, we reveal the observation of superfluid motion of a quantum fluid of polaritons created by a laser in a semiconductor microcavity.

Bound electron-hole particles, known as excitons, are fascinating objects in semiconductor nanostructures. In a quantum well with a thickness of the order of a few nanometres, the external motion of the excitons is quantized in the direction perpendicular to the well, while it is free within the plane of the well. When the quantum well is placed in a high finesse microcavity, the strong coupling regime between excitons and light is easily reached[12], giving rise to exciton-photon mixed quasi-particles called polaritons, which are an interesting kind of two-dimensional composite bosons. Thanks to their sharp dispersion, polaritons have a small effective mass (of the order of $10^{-5}$ times the free electron mass) that enables the building of many-body quantum coherent effects, like Bose Einstein condensation[13, 14], at a lattice temperature of a few kelvins.

In order to probe the superfluidity in the polariton system we study the perturbation that is produced in an optically created moving polariton fluid when a static defect of the microcavity is present in the flow path, as proposed in Refs. 15, 16. This procedure is a direct application to the polariton system of the standard Landau criterion of superfluidity[5] originally developed for liquid Helium and recently applied to demonstrate superfluidity of atomic BECs[17, 18].

To explore the quantum fluid regime a complete control of three key-parameters is needed: the in-plane momentum of polaritons (i.e., the polariton flow velocity), the oscillation frequency of the polariton field, and its density. In this respect polaritons constitute an ideal system from the experimental point of view. Due to their excitonic component they are subject to binary interactions, therefore, the associated interaction energy within the polariton fluid can be controlled through the polariton density, which in turn is changed in a precise way by adjusting the incident laser power. Their partially photonic character also enables the creation of polariton fluids with a well defined oscillation frequency, which is the frequency $\omega_p$ of the excitation laser, and with a well defined linear momentum $k_p$, by choosing the angle of incidence $\theta_p$ [$k_p = (\omega_p/c) \sin \theta_p$, where $c$ is the light speed]. The possibility of controlling the polariton fluid oscillation frequency is in stark contrast with equilibrium systems, like atomic condensates, where the oscillation frequency of the condensate is fixed by the equation of state relating the chemical potential to the particle density[3]. Indeed, polariton gases constitute a strongly non-equilibrium system: due to their relatively short lifetime (of the order of few picoseconds), the steady state of an excited microcavity system results from the interplay between the pumping rate and the radiative as well as non-radiative losses. This feature results in much wider possibilities in the structure of the system's spectrum of elementary excitations than in the equilibrium case[15, 16, 19-22].

In our experiment a polariton fluid is excited in a microcavity sample, cooled at 5 K, with a circularly polarized beam from a frequency stabilized, single mode continuous wave Titanium:Sapphire laser. The laser field continuously replenishes the escaping polaritons in the fluid. The beam is focused onto the sample in a spot of ~100 μm in diameter with angles of incidence between 2.6º and 4.0° (Fig. 1a). The wavelength of the pump laser is around 836 nm, close to resonance with the lower polariton branch (LPB). The image of the surface of the sample (near field emission) and of the far field in transmission configuration are simultaneously recorded on two different high resolution CCD cameras. With the use of a spectrometer, and at low power off resonance excitation, the characteristic parabolic lower polariton dispersion can be observed, as shown in Fig. 1b.

In order to study the propagation properties of the injected polariton fluid, the centre of the excitation spot is placed on top of a natural point-like defect present in the sample. Defects of different sizes and shapes appear naturally in the growth process of microcavity sample. At low excitation power and quasi resonant excitation of the LPB, polariton-polariton interactions are negligible: in the near-field (real space) images, the coherent polariton gas created by the laser is scattered by the defect and generates a series of parabolic-like wavefronts around the defect, propagating away from it, mostly in the upstream direction (Figs. 2c-I, and 3b-I). They result from the interference of an incident polariton plane-wave with a cylindrical wave produced by the scattering on the defect. In momentum space, polariton scattering gives rise to the well-known Rayleigh ring[23] that is observed in the far-field images (Figs. 2c-IV, 3b-IV).

As the laser intensity is augmented, polariton-polariton interactions increase resulting in the single-polariton dispersion curves being shifted towards higher energies (blue shift due to the repulsive interactions) and also becoming strongly distorted as a consequence of collective Bogoliubov-like manybody interactions[15, 16]. In a simplified picture, from parabolic (Fig. 1c) the dispersion is predicted to become linear in some $k$ vector range with a discontinuity of its slope in the vicinity of the pump wave vector $k_p$ [see Fig. 1d and Refs. 15, 16]. Under these conditions, a sound velocity can be attributed to the polariton fluid, being given by:

$$c_s = \sqrt{\hbar g |\psi|^2 / m} \qquad (1),$$

where $g$ is the polariton-polariton coupling strength, $|\psi|^2$ is the polariton density and $m$ the effective mass of the lower polariton branch. If the excitation density of the polariton fluid and its flow velocity $v_p$ (given by $v_p = \hbar k_p / m$) are controlled in such a way to have $v_p < c_s$, then the Landau criterion for superfluidity is satisfied, as shown in Ref. 16. In such a case, since no states are any longer available for scattering at the frequency of the driving polariton field (see Fig. 1d), the polariton scattering from the defect is inhibited and the polariton fluid is able to flow without friction. This situation is observed in Fig. 2 where the real (c-III) and momentum (c-VI) space images of the polariton fluid in the presence of a ~4 μm diameter defect are shown for a pump angle of incidence of 2.6°, corresponding to a low in-plane momentum of $k_\parallel$ = -0.337 μm$^{-1}$ ($v_p$ = 6.4x10$^5$ m·s$^{-1}$, point A in Fig. 1b). Simulations based on the solution of polariton non-equilibrium Gross-Pitaevskii equations (see Ref 16) are shown in Fig. 2d-III and 2d-VI. The calculations have been performed without the use of any fitting parameter apart from the size and depth of the defect. While, at low excitation density (Fig. 2c-I, c-IV, d-I, d-IV), the fluid presents parabolic density wavefronts in real space, and a scattering ring in momentum space as mentioned above, at higher excitation density, the scattering ring collapses (Fig. 2c-V-VI, d-V-VI) showing that any scattering of the polariton fluid by the defect is inhibited, and that frictionless flow is eventually attained. In real space (Figs. 2c-III, 2d-III), a complete suppression of the density modulation is observed. In all these figures, one can see an excellent agreement between the observed effects and the theory.

The collapse of the scattering ring in the superfluid regime is clearly summarized in Fig. 2b, where the ratio between the polaritons scattered to a constant area in momentum space (dashed yellow square in Fig. 2c-IV), and the total polariton density is plotted as a function of the excitation density: when the superfluid regime is obtained, the scattered light drops by a factor of 4. Note that this factor is here limited by the finite size of the excitation spot and is expected to attain much higher values if larger pump spots were used.

It is important to stress that the evidence of superfluidity presented here is substantially different from what was observed in Ref. 11, which shows the dispersionless propagation of a polariton bullet even when crossing a defect. Although these results strongly suggest a superfluid character, the propagation of the sizeable polariton bullet on top of a homogeneous fluid under parametric scattering conditions can not be described in terms of the Bogoliubov theory of a weakly perturbed fluid. In contrast, the experimental study reported in the present paper constitutes a direct application of the Landau criterion and allows a clear demonstration of superfluidity in a quasi-homogeneous polariton fluid. It has also been suggested to exploit the Landau criterion to investigate photonic superfluidity in non linear optical systems[24-26] but no experimental studies have been yet reported.

The qualitative shape of the perturbation created in the fluid by the defect is studied in Fig.3 in a different regime. For this purpose, a polariton fluid with a higher momentum is created using a laser beam with a larger incidence angle of 4.0º ($k_p = -0.521$ μm$^{-1}$, point B in Fig.1b). This allows us to enter a regime where the Bogoliubov dispersion of collective excitations has a sound-like nature with a sound speed which is now lower than the flow speed ($v_p > c_s$). The generalised Landau condition for superfluidity is not fulfilled and the defect is able to generate collective Bogoliubov excitations in the fluid. These manifest as a Cerenkov-like density modulation pattern with characteristic straight wavefronts in the real space images (b-II-III) and as a strongly reshaped Rayleigh scattering ring in the far field emission pattern[15, 16] (b-V-VI). A similar density pattern was recently observed in atomic BECs propagating against the optical potential of a localised defect at a super-sonic velocity[27]. At lower pump power (b-I, IV), the Bogoliubov excitations go back to single-particle ones, and the parabolic-shaped modulation corresponding to the standard Rayleigh scattering ring is recovered as in Fig.2c-I, IV. The calculated images also reproduce these observations as shown in (c-I-III) and (c-IV-VI): again, the agreement with the experimental observations is excellent.

Let us note that the observation of the linear Cerenkov-like wavefronts in Fig. 3b-II indicates the existence of a well defined sound speed in the system. The angle of aperture $\phi$ of the wake profile enables the precise measurement of the sound speed in the system, given by $\sin(\phi) = c_s/v_p$. In the conditions of Fig. 3b-II we find $c_s = 8.1 \times 10^5$ m·s$^{-1}$. With the use of Eq. 1 the polariton-polariton coupling constant *g* can be estimated, being of the order of 0.01 meV μm$^2$, which is the value of used in the simulations.

In conclusion, we have reported an experimental demonstration of superfluidity in a quantum fluid of exciton-polaritons flowing in a semiconductor microcavity. Superfluidity as defined by the Landau's criterion has been demonstrated by varying both the flow velocity and the density of polaritons. By varying the pump intensity, we have observed that the system goes from a non-superfluid regime in which a static defect creates a substantial perturbation in the moving fluid to a superfluid one, in which the polariton flow is no longer affected by the defect. In a supersonic regime, superfluid propagation is replaced by the appearance of a Cerenkov-like perturbation produced by the defect, in agreement with theoretical predictions and detailed calculations. Our observations pave the way towards the investigation of a rich variety of quantum fluid effects associated to the non-equilibrium nature of the microcavity polariton system.

**Acknowledgements**

This work was partially supported by the *Ile de France* programme IFRAF. A.A. and S.P. were funded by the *Agence Nationale pour la Recherche*, A.B. is a member of the *Institut Universitaire de France*. IC acknowledges financial support from the Italian MIUR and the EuroQUAM-FerMix program.

Correspondence and requests for materials should be addressed to A.A or A.B.

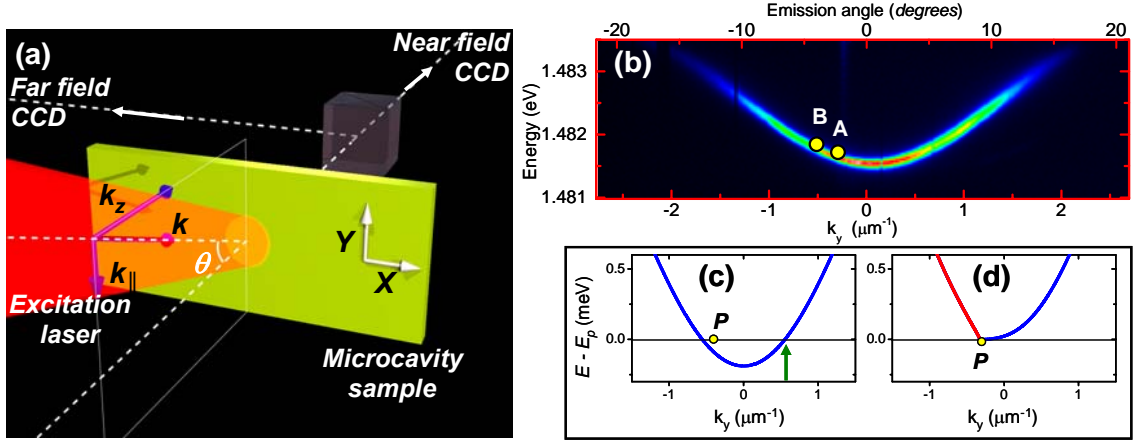

**Fig.1.** (a) Overview of the experimental excitation and detection conditions. (b) Lower polariton branch dispersion in the linear regime as observed after non-resonant excitation. Points A and B denote the excitation energy and momentum corresponding to the results shown in Figs. 2 and 3, respectively. (c) Analytically calculated spectrum of excitations under low power resonant pumping at the point indicated by the yellow dot (*P*). Injected polaritons can elastically scatter to same energy states, as those indicated by the green arrow. $E_p$ refers to the energy of the pump beam. (d) Analytically calculated spectrum of excitations under strong resonant pumping under the conditions of superfluidity –Figs. 2c-III, 2c-VI and 2d-III, 2d-VI–, where the Landau criterion is fulfilled and injected polaritons cannot scatter due to the absence of available final states at the energy of the pump. The red section evidences the strongly modified linear shape due to polariton-polariton interaction. Figures (c), (d) have been calculated for a spatially homogeneous system with the same parameters as those at the centre of the spot in Fig. 2.

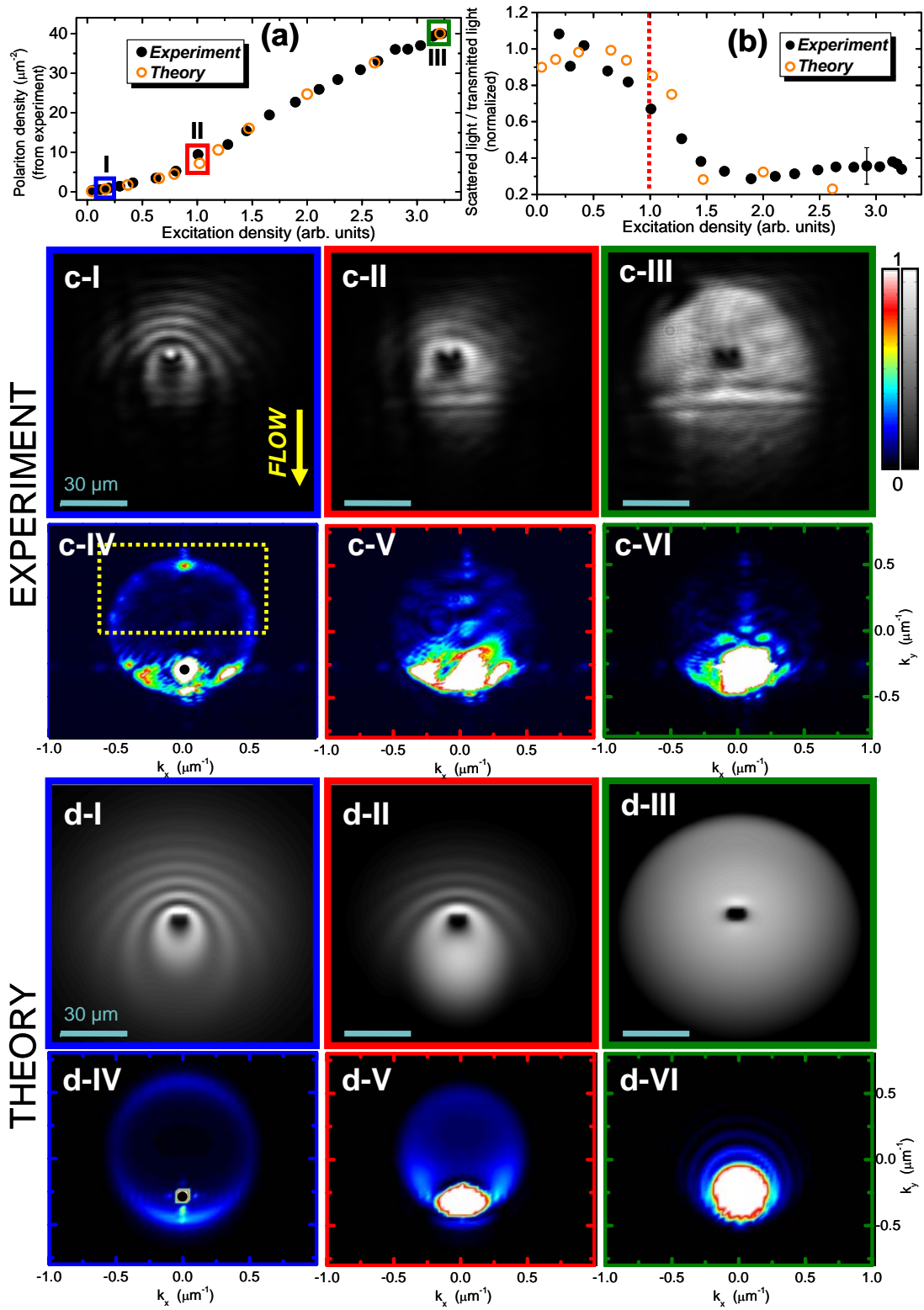

**Fig. 2. Superfluid regime.** Observation of polariton fluids created with a low in plane momentum of -0.337 μm⁻¹ (excitation angle 2.6°), and an excitation laser blue-detuning

of 0.1 meV with respect to the low density polariton dispersion (point A in Fig. 1a). (a) Experimentally observed (solid points) and calculated (open points) transmitted intensity (proportional to the polariton density) as a function of the excitation power. Panels c-I-III (c-IV-VI) depict the experimental near field (far field, i.e., momentum space) images of the excitation spot around a defect present in the sample, for the corresponding excitation densities marked in (a) with coloured squares. At low power (c-I, blue) when polariton-polariton interactions are negligible, the polariton fluid scatters on the defect giving rise to characteristic parabolic wavefronts, and a corresponding elastic scattering ring (c-IV). At high powers, the emission patterns are significantly affected by polariton-polariton interactions (c-II, red) and eventually show the onset of a superfluid regime characterized by an undisturbed flow around the defect (c-III, green). In momentum space, the approach and eventual onset of a superfluid regime is evidenced by the shrinkage (c-V) and then collapse (c-VI) of the scattering ring. Panels (d) show the corresponding calculated images. The black solid dot in (c-IV) and (d-IV) indicates the momentum coordinates of the excitation beam. All the far-field images are normalized to the total transmitted intensity and saturated (white area) to improve the visibility of the scattered emission. The photonic defect is repulsive (i.e., the value of the photonic potential inside the defect is greater than outside) and has a depth of ~1 meV and a size of ~4 μm, while the excitation spot extends over an area of 84 μm in diameter (full width at $1/e$). (b) Shows the relative scattered polariton intensity as a function of excitation density, as calculated (open points) and measured experimentally (solid points) in an area in momentum space indicated by the yellow square in (c-IV). At the onset of the superfluid regime (red line) the relative amount of scattered polaritons drops by a factor of 4, as the Rayleigh scattering ring collapses.

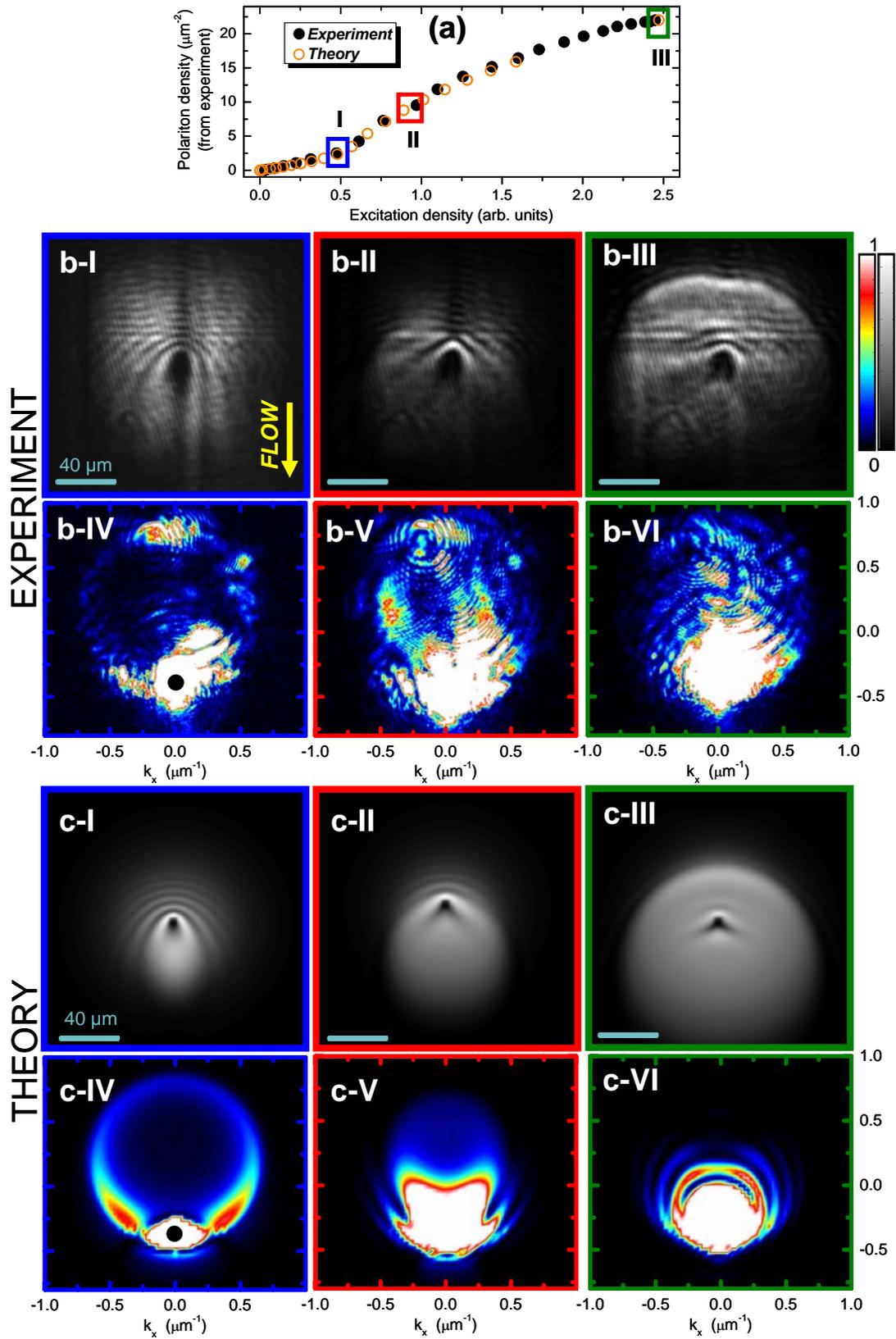

**Fig. 3. Cerenkov regime.** Observation of a polariton fluid created with an in plane

momentum of -0.521 $\mu m^{-1}$ (angle of incidence of 4.0º) and an excitation laser blue-detuning of 0.11 meV with respect to the low density polariton dispersion -point B in Fig. 1b-. (a) Experimentally observed (solid points) and calculated (open points) polariton density as a function of the excitation power. Panels b-I-III (b-IV-VI) depict the experimental near field (far field, i.e., momentum space) images of the excitation spot around a repulsive photonic defect present in the sample, for the corresponding excitation densities marked in (a) with coloured squares. At low power, in the absence of polariton-polariton interactions, the polariton emission is characterized by parabolic wavefronts around the defect in real space (b-I, blue) and by the appearance of a Rayleigh elastic scattering ring in momentum space (b-VI). As the excitation density is increased, the onset of polariton-polariton interactions leads the system to the Cerenkov regime characterized by the linear wavefronts in real space around the defect (b-II, red). Under these conditions the polariton fluid velocity is higher than its sound velocity. In momentum space a strong modification of the scattering ring is observed (b-V). At higher excitation densities the Cerenkov-like regime is maintained (b-III, b-VI). Panels (c) show the corresponding calculated images. The solid dot in (b-IV) and (c-IV) indicates the momentum coordinates of the excitation beam. The dotted red lines are guides to the eye evidencing the shrinkage of the scattering ring. All the far-field images are normalized to the total transmitted intensity. The defect has a height of ~1 meV and a size of ~5 $\mu$m.